\documentclass[aps,prl,twocolumn,preprintnumbers]{revtex4-1}

\usepackage{amsmath}
\usepackage{hyperref}
\usepackage{graphicx}
\usepackage{color}

\def\beq{\begin{equation}}
\def\eeq#1{\label{#1}\end{equation}}
\def\eeqn{\end{equation}}

\def\beqa{\begin{eqnarray}}
\def\eeqa#1{\label{#1}\end{eqnarray}}
\def\eeqan{\end{eqnarray}}
\def\CR{\nonumber \\ }

\def\eqref#1{(\ref{#1})}

\def\L{{\cal L}}
\def\O{{\cal O}}

\def\sw{s_\theta}
\def\cw{c_\theta}
\def\VCKM{V_{\text{CKM}}}

\def\dgz{\delta g_{1z}}
\def\dkz{\delta\kappa_z}
\def\dka{\delta\kappa_\gamma}
\def\lz{\lambda_z}
\def\la{\lambda_\gamma}

\def\dgLZf{\delta g_L^{Zf}}
\def\dgLZfp{\delta g_L^{Zf'}}
\def\dgRZf{\delta g_R^{Zf}}
\def\dgLRZf{\delta g_{L/R}^{Zf}}
\def\dgLWF{\delta g_L^{WF}}
\def\dgLWq{\delta g_L^{Wq}}
\def\dgLWl{\delta g_L^{Wl}}
\def\dgLZu{\delta g_L^{Zu}}
\def\dgLZd{\delta g_L^{Zd}}
\def\dgRZu{\delta g_R^{Zu}}
\def\dgRZd{\delta g_R^{Zd}}
\def\dgLZv{\delta g_L^{Z\nu}}
\def\dgLZe{\delta g_L^{Ze}}
\def\dgRZe{\delta g_R^{Ze}}

\allowdisplaybreaks

\begin{document}

\preprint{MCTP-16-24}
\preprint{DESY-16-189}

\title{Time to Go Beyond Triple-Gauge-Boson-Coupling Interpretation \\ of $W$ Pair Production}


\author{Zhengkang Zhang}
\affiliation{Michigan Center for Theoretical Physics (MCTP), University of Michigan, Ann Arbor, Michigan 48109, USA}
\affiliation{Deutsches Elektronen-Synchrotron (DESY), 22607 Hamburg, Germany}



\begin{abstract}
$W$ boson pair production processes at $e^+e^-$ and $pp$ colliders have been conventionally interpreted as measurements of $WWZ$ and $WW\gamma$ triple gauge couplings (TGCs). Such interpretation is based on the assumption that new physics effects other than anomalous TGCs are negligible. While this ``TGC dominance assumption'' was well-motivated and useful at LEP2 thanks to precision electroweak constraints, it is already challenged by recent LHC data. In fact, contributions from anomalous $Z$ boson couplings that are allowed by electroweak precision data but neglected in LHC analyses, being enhanced at high energy, can even dominate over those from the anomalous TGCs considered. This limits the generality of the anomalous TGC constraints derived in current analyses, and necessitates extension of the analysis framework and change of physics interpretation. The issue will persist as we continue to explore the high energy frontier. We clarify and analyze the situation in the effective field theory framework, which provides a useful organizing principle for understanding Standard Model deviations in the high energy regime.
\end{abstract}


\maketitle


{\it Introduction.}---The nonabelian nature of the Standard Model (SM) gauge groups has the crucial consequence of gauge boson self-interactions. In the electroweak sector, the structure of $WWZ$ and $WW\gamma$ triple gauge couplings (TGCs) follows from $SU(2)_L\times U(1)_Y$ gauge invariance and the pattern of its spontaneous breaking. Heavy new physics beyond the SM may leave footprints on the low-energy effective theory in the form of anomalous TGCs. Conventionally, these effects are parameterized by the following Lagrangian \cite{TGC87}, 
\beqa
&&\L_{\text{TGC}} =\CR
&&\quad ig \,\Bigl\{ (W^+_{\mu\nu}W^{-\mu}-W^-_{\mu\nu}W^{+\mu}) \bigl[ (1+\dgz)\, \cw Z^\nu +\sw A^\nu \bigr] \CR
&&\quad\,\, +\frac{1}{2} W^+_{[\mu,}W^-_{\nu]} \bigl[ (1+\dkz) \,\cw Z^{\mu\nu} +(1 +\dka)\, \sw A^{\mu\nu} \bigr] \CR
&&\quad\,\, +\frac{1}{m_W^2} W^{+\nu}_\mu W^{-\rho}_\nu (\lz\cw Z_\rho ^{\,\,\,\mu}+\la\sw A_\rho^{\,\,\,\mu}) \Bigr\},
\eeqa{eq:LTGC}
assuming $CP$ conservation. Here $W^\pm_{\mu\nu}=\partial_\mu W^\pm_\nu-\partial_\nu W^\pm_\mu$, $W^+_{[\mu,}W^-_{\nu]}=W^+_\mu W^-_\nu-W^+_\nu W^-_\mu$, $g$ is the $SU(2)_L$ gauge coupling, and $\sw$ ($\cw$) denotes the sine (cosine) of the weak mixing angle. The anomalous TGC parameters $\dgz,\dkz,\dka,\lz,\la$, which vanish in the SM, have been intensively studied in search of evidence for new physics. LEP2 measurements of $W$ pair (and to a lesser extent also single $W$) production were able to confirm SM predictions and constrain the anomalous TGCs at the $\lesssim$\,10\% level \cite{LEP2,FalkowskiRiva}. Recent years have seen renewed interest in TGC studies, motivated by progress on LHC electroweak measurements as well as connection with Higgs physics \cite{TGCfromHiggs,FalkowskiReview,TGCglobal}. Impressively, with several diboson measurements at 7 and 8\,TeV combined, LHC has already exceeded LEP2 in setting limits on anomalous TGCs \cite{GHlegacy}. The $WW$ \cite{ATLASWW,CMSWW} and $WZ$ \cite{ATLASWZ,CMSWZ} channels played a dominant role in this achievement. Prospects of future facilities have also been discussed, with numbers as small as $10^{-4}$--$10^{-3}$ quoted for anomalous TGC sensitivities \cite{ILCTDR,TGCCEPC}, showing great potential of uncovering new physics beyond the SM in electroweak interactions.

In previous TGC analyses, it is often assumed that Eq.~\eqref{eq:LTGC} encodes {\it all} the relevant beyond-SM effects on the observables under study. This assumption, which we shall call the ``TGC dominance assumption,'' is obviously not satisfied for arbitrary new physics scenarios. Nevertheless, it is well-motivated and useful if other possible deformations of the SM are experimentally constrained to be small. Whether the latter is the case should be carefully assessed to give meaning to TGC studies.

To do so, we consider the most general SM deformations due to decoupled new physics at a high scale $\Lambda$, which can be captured by the SM effective field theory (EFT) at experimentally accessible energies, assumed to be much below $\Lambda$. Generically, assuming lepton number conservation up to $\Lambda$, leading corrections to the SM Lagrangian arise from dimension-six effective operators,
\beq
\L_{\text{SMEFT}} = \L_{\text{SM}} + \sum_i \frac{C_i}{v^2}\O_i +\dots,
\eeqn
with $C_i\sim\O(\frac{v^2}{\Lambda^2})$ up to model-dependent coupling or loop factors. 
In this framework, search of SM deviations becomes a global analysis program, with experimental input on different processes probing various directions of the $\{C_i\}$ parameter space \cite{UltimateFit,ESYfit,BBTfit}. Conventional TGC analyses via e.g.\ $W$ pair production are usually interpreted as constraining the linear combinations of $C_i$ giving rise to anomalous TGCs. This is based on the TGC dominance assumption, which asserts that other combinations of $C_i$ affecting the observables under study can be effectively set to zero. It is commonly believed that the latter combinations are well constrained by electroweak precision data (EWPD) and can hardly have any impact.

It is the purpose of this letter to revisit the TGC dominance assumption and assess its validity in light of recent improvements on TGC constraints from LHC data. We focus on the $WW$ channel given its relevance at both $e^+e^-$ and $pp$ colliders. 
There have been claims in the previous literature that the TGC dominance assumption is valid in the case of $e^+e^-\to W^+W^-$ at LEP2~\cite{UltimateFit,FalkowskiRiva}, and we will verify them explicitly. On the other hand, as we will see, 
this assumption is no longer supported by EWPD when analyzing recent LHC results. A key observation is that some of the neglected effects, even though constrained by $Z$ pole data, are enhanced at higher energies and become nonnegligible compared with the anomalous TGC effects under study. The situation calls for extension of the TGC analysis framework to allow for more general interpretations of experimental results. Further, from the SMEFT point of view, as we continue to explore the high energy frontier, it will be perhaps more useful to organize our knowledge of effective operators in terms of their high energy behaviors, rather than the anomalous couplings they induce as is conventionally done.

\vspace{3pt}

{\it Effective operators and anomalous couplings.}---We start by reviewing the theoretical framework in order to precisely formulate the TGC dominance assumption. We shall be guided by the SMEFT at dimension-six level to identify potentially important beyond-SM effects in addition to anomalous TGCs. In the Warsaw basis \cite{WarsawBasis}, which we adopt here for concreteness, the following operators contribute to $f\bar f\to W^+W^-$ at tree level:
\beqa
&& \O_{HWB} = H^\dagger\sigma^a H W^a_{\mu\nu}B^{\mu\nu},\quad
\O_{HD} = |H^\dagger (D_\mu H)|^2,\CR
&& \O_{3W}=\epsilon^{abc}W_\mu^{a\nu}W_\nu^{b\rho}W_\rho^{c\mu}, \quad
\bigl[\O_{ll}\bigr]_{ijkn} = (\bar l_i\gamma_\mu l_j)(\bar l_k\gamma^\mu l_n), \CR
&& \bigl[\O_{HF}^{(3)}\bigr]_{ij}=i\bigl(H^\dagger\sigma^a (D_\mu H) -(D_\mu H^\dagger)\sigma^a H\bigr) (\bar F_i\gamma^\mu\sigma^a F_j), \CR
&& \bigl[\O_{HF}^{(1)}\bigr]_{ij} = i\bigl(H^\dagger (D_\mu H) -(D_\mu H^\dagger) H\bigr) (\bar F_i\gamma^\mu F_j), \CR
&& \bigl[\O_{Hf}\bigr]_{ij} = i\bigl(H^\dagger (D_\mu H) -(D_\mu H^\dagger)H\bigr) (\bar f_i\gamma^\mu f_j),
\eeqa{eq:OWarsaw}
where $F$, $f$ denote $SU(2)_L$ doublet and singlet fields, respectively, and $i,j$ are generation indices. We assume minimal flavor violation \cite{MFV} for simplicity, and neglect operators whose coefficients are suppressed by SM Yukawa couplings. An alternative dimension-six operator basis commonly used in the literature is discussed in the Supplemental Material~\cite{SM}.

One can work out the anomalous couplings induced by the dimension-six operators in Eq.~\eqref{eq:OWarsaw}. To avoid ambiguities associated with field and parameter redefinitions, we follow \cite{BSMprimaries,HiggsBasis} and define anomalous couplings with respect to mass eigenstate fields in unitary gauge with canonically normalized kinetic terms, after SM parameters have been properly redefined such that the conventional input observables $m_Z$, $G_F$, $\alpha$, etc.\ are not shifted (see \cite{ETUT} for connection with the oblique parameters formalism \cite{PeskinTakeuchi,BeyondSTU,STWY}). In this framework, $f\bar f\to W^+W^-$ can receive new physics contributions from: 
{\it i}) anomalous TGCs defined in Eq.~\eqref{eq:LTGC}; 
{\it ii}) $W$ boson mass shift
\beq
\L_{m_W} = (1+\delta_m)^2 \frac{g^2v^2}{4}W^+_\mu W^{-\mu};
\eeqn
and {\it iii}) $Zff$ and $Wff'$ vertex corrections (with $f'$ being the $SU(2)_L$ partner of $f$)
\beqa
\L_{\text{vertex}} &=& \sum_f \frac{g}{\cw}\bigl((T^3_f-Q_f\sw^2)\delta_{ij} +\bigl[\dgLRZf\bigr]_{ij}\bigr) Z_\mu\bar f_i\gamma^\mu f_j \CR
&& +\frac{g}{\sqrt{2}}\Bigl[\bigl(\delta_{ij}+\bigl[\dgLWq\bigr]_{ij}\bigr)W^+_\mu \bar u_{Li}\gamma^\mu (\VCKM d_L)_j \CR
&&+\bigl(\delta_{ij}+\bigl[\dgLWl\bigr]_{ij}\bigr)W^+_\mu \bar \nu_i\gamma^\mu e_{Lj}+\text{h.c.}\Bigr],
\eeqa{eq:Lvertex}
where $f$ now runs over mass eigenstates $\nu_L$, $e_{L,R}$, $u_{L,R}$, $d_{L,R}$, 
and 
generation indices $i,j$ are summed over. These anomalous couplings are not all independent. In particular, anomalous TGCs satisfy the well-known relations,
\beq
\dkz = \dgz -\frac{\sw^2}{\cw^2}\dka,\quad \lz=\la,
\eeqn
while $Zff$ and $Wff'$ vertex corrections are also related,
\beq
\dgLWq = \dgLZu-\dgLZd,\quad \dgLWl = \dgLZv-\dgLZe.
\eeqn
Therefore, there are 5(6) independent anomalous couplings contributing to $f\bar f\to W^+W^-$ with right-handed (left-handed) incoming fermion: $\dgz$, $\dka$, $\la$, $\delta_m$, plus $\dgRZf$ ($\dgLZf$ and $\dgLZfp$). In particular, $e^+e^-\to W^+W^-$ at LEP2 involves 7 independent anomalous couplings
\beq
\bigl\{\dgz, \dka, \la, \dgLZe, \dgLZv, \dgRZe, \delta_m\bigr\},
\eeq{eq:ACeeww}
while $pp\to W^+W^-$ at the LHC involves 8 when only first-generation quarks are considered in the initial state
\beq
\bigl\{\dgz, \dka, \la, \dgLZu, \dgRZu, \dgLZd, \dgRZd, \delta_m\bigr\},
\eeq{eq:ACppww}
with generation indices $i,j=1$ implicit. 

The dictionary between effective operator coefficients and the anomalous couplings listed above reads
\beqa
&& \dgz = \frac{1}{\cw^2-\sw^2} \Bigl(-\frac{\sw}{\cw}C_{HWB}-\frac{1}{4}C_{HD}-\delta v\Bigr), 
\CR
&& \dka = \frac{\cw}{\sw}C_{HWB}, \quad \la = -\frac{3}{2}g\,C_{3W}, 
\CR
&& \bigl[\dgLZf\bigr]_{ij} = T^3_f \bigl[C_{HF}^{(3)}\bigr]_{ij} -\frac{1}{2}\bigl[C_{HF}^{(1)}\bigr]_{ij} -\biggl[ Q_f \frac{\cw\sw}{\cw^2-\sw^2} C_{HWB} \CR
&&\qquad\qquad +\Bigl(T^3_f+Q_f\frac{\sw^2}{\cw^2-\sw^2}\Bigr) \Bigl(\frac{1}{4}C_{HD}+\delta v\Bigr) \biggr] \delta_{ij}, 
\CR
&& \bigl[\dgRZf\bigr]_{ij} = -\frac{1}{2}\bigl[C_{Hf}\bigr]_{ij} -Q_f \biggl[ \frac{\cw\sw}{\cw^2-\sw^2} C_{HWB} \CR
&&\qquad\qquad +\frac{\sw^2}{\cw^2-\sw^2}\Bigl(\frac{1}{4}C_{HD}+\delta v\Bigr) \biggr] \delta_{ij},
\CR
&&\delta_m = -\frac{1}{\cw^2-\sw^2}\Bigl(\cw\sw C_{HWB} +\frac{1}{4}\cw^2C_{HD}+\sw^2\delta v\Bigr),
\eeqa{eq:ACWarsaw}
where $F$ denotes the $SU(2)_L$ doublet containing $f_L$, and $\delta v \equiv \frac{1}{2} \bigl([C_{Hl}^{(3)}]_{11}+[C_{Hl}^{(3)}]_{22}\bigr) -\frac{1}{4} \bigl([C_{ll}]_{1221}+[C_{ll}]_{2112}\bigr)$.

With the discussion above, it should be clear that, as far as the dimension-six SMEFT is concerned, the TGC dominance assumption corresponds to keeping only the subset $\{\dgz, \dka, \la\}$ of anomalous couplings in Eqs.~\eqref{eq:ACeeww} and~\eqref{eq:ACppww}. We see from Eq.~\eqref{eq:ACWarsaw} that, once the operators inducing $\dgz$, $\dka$ are turned on, one then has to adjust $C_{HF}^{(3,1)}$, $C_{Hf}$ to ensure that vertex corrections vanish.

\vspace{3pt}

{\it Triple gauge coupling measurements: from LEP2 to LHC.}---Now we make a first attempt to assess the validity of the TGC dominance assumption in $W$ pair production processes. For illustration, we will allow each of the additional anomalous couplings,
\beq
\bigl\{\dgLZe, \dgLZv, \dgRZe, \dgLZu, \dgRZu, \dgLZd, \dgRZd, \delta_m\bigr\}
\eeq{eq:ACpole}
to be maximal within the $2\sigma$ intervals in Eq.~(40) of \cite{FalkowskiRiva} and Eq.~(4.4) of \cite{flavorful}, which are derived from EWPD assuming flavor universality, and see how much correction they can induce on some representative observables. This is to be compared with contributions from anomalous TGCs being considered in conventional TGC analyses, as well as experimental uncertainties.

\begin{figure}[tbp]
\centering
\includegraphics[width=3.3in]{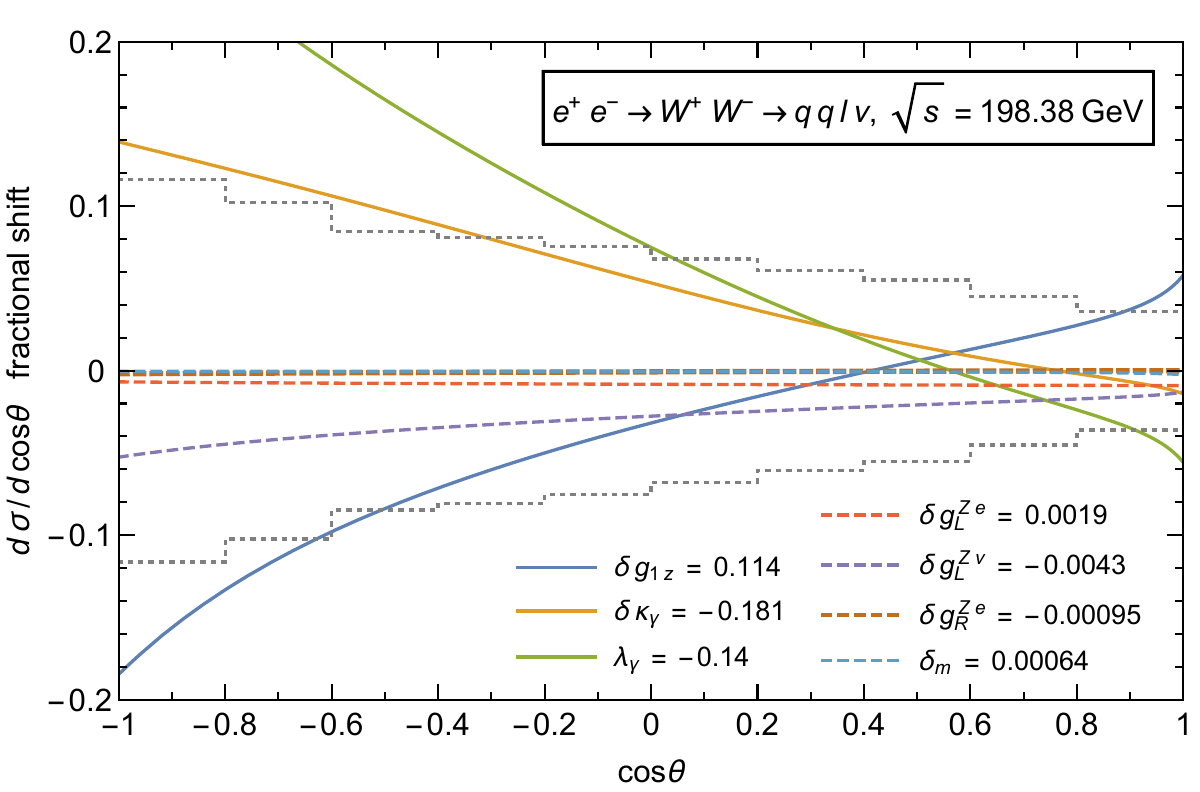}
\caption{Fractional shift in LEP2 $e^+e^-\to W^+W^-\to qq\ell\nu$ differential cross section induced by each of the anomalous couplings in Eq.~\eqref{eq:ACeeww}, compared with experimental uncertainties (gray dotted) reported in \cite{LEP2}. Assuming lepton flavor universality, effects of the anomalous TGCs being constrained (solid) \cite{LEP2002} are seen to dominate over those of $Zff$ vertex and $W$ mass corrections (dashed), even when the latter are set to maximum values allowed by EWPD \cite{FalkowskiRiva,flavorful}, providing justification for the conventional TGC analysis procedure.
\label{fig:LEP}}
\end{figure}

TGC analyses at LEP2 made use of $e^+e^-\to W^+W^-$ measurements with unpolarized $e^+e^-$ beams at center-of-mass energies up to 209\,GeV. We consider as an example observable $\frac{d\sigma}{d\cos\theta}(e^+e^-\to W^+W^-\to qq\ell\nu)$ with $\theta$ being the angle between $W^-$ and $e^-$ momenta and $\ell=e,\mu$ (either sign), at $\sqrt{s}=198.38$\,GeV. This is reported for 10 bins of $\cos\theta$ in Table~5.6 of \cite{LEP2} based on data from 194 to 204\,GeV, with a luminosity-weighted average of 198.38\,GeV. Fig.~\ref{fig:LEP} shows the fractional shift in $\frac{d\sigma}{d\cos\theta}$ with respect to the SM, calculated at tree level, when each of the anomalous couplings in Eq.~\eqref{eq:ACeeww} is turned on individually, along with experimental uncertainties (gray dotted). Contributions from $\dgLZu$, $\dgLZd$ via $W$ branching ratio modifications are within $\pm0.005$ and not shown. Numerical values chosen for the anomalous TGCs reflect the level of LEP2 constraints --- they correspond to maximal deviations from zero allowed by the LEP2 three-parameter fit (95\% C.L.\ intervals in Table~11.7 of \cite{LEP2002}). It is seen that possible contributions from vertex and $W$ mass corrections as allowed by EWPD are indeed well beyond experimental sensitivity, thus providing justification for the conventional TGC analysis procedure (though the situation may be more subtle when theoretical errors from EFT calculations are considered \cite{BBTfit}).

\begin{figure}[tbp]
\centering
\includegraphics[width=3.3in]{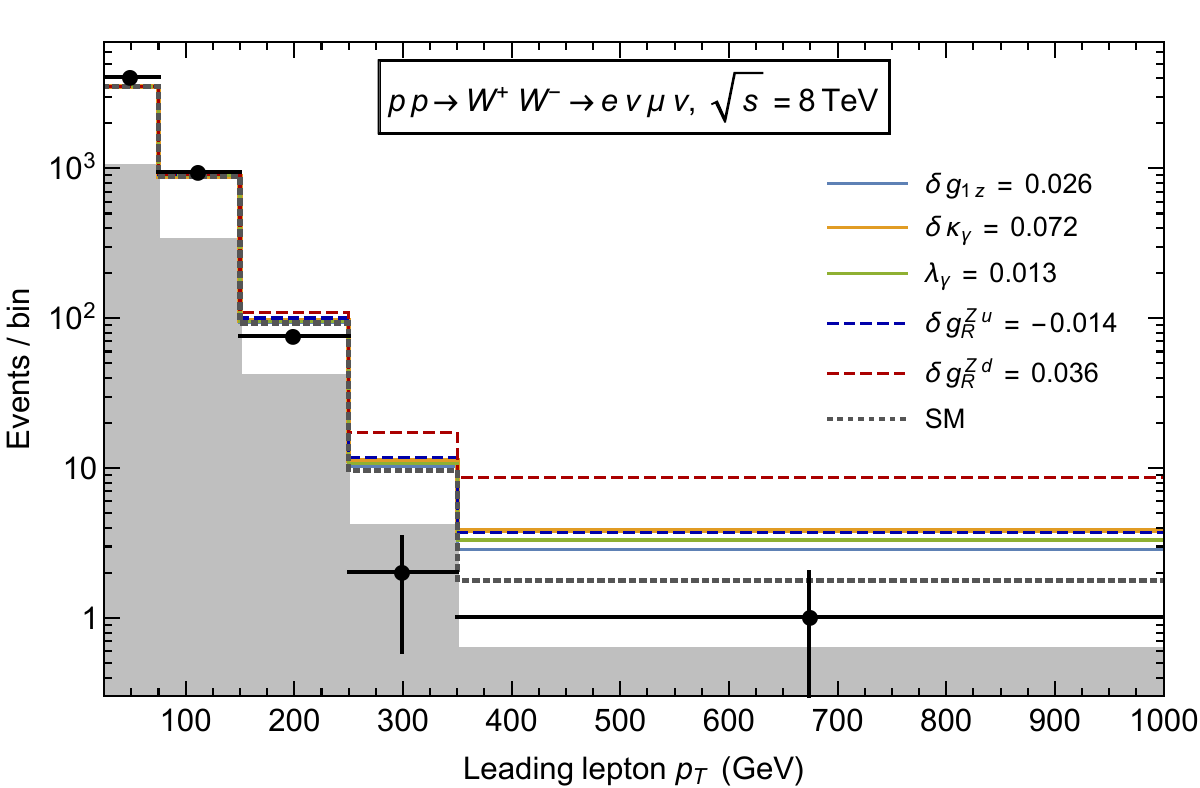}
\caption{Leading lepton $p_T$ distribution of 8\;TeV LHC $W^+W^-$ events  in the $e\mu$ channel when each anomalous coupling is turned on individually, compared with experimental data (dots with error bars) and SM predictions (gray dotted). The latter, together with non-$WW$ backgrounds (gray shaded), are taken from \cite{ATLASWW}. Effects of anomalous TGCs being considered in recent TGC fits (solid) are clearly {\it not} dominant over those of $\dgRZu$, $\dgRZd$ (dashed) consistent with EWPD, calling for extension of the conventional TGC analysis procedure.
\label{fig:LHC}}
\end{figure}

At the LHC, anomalous TGC constraints greatly benefit from the $W^+W^-\to e^\pm\mu^\mp\nu\nu$ channel. For illustration, we follow the ATLAS 8\;TeV analysis \cite{ATLASWW}, and numerically calculate the leading lepton $p_T$ distribution. Technical details of the calculation can be found in the Supplemental Material~\cite{SM}. Our results are shown in Fig.~\ref{fig:LHC}, with values of anomalous TGCs chosen at the 95\% C.L.\ upper limits from the recent TGC fit \cite{GHlegacy}, which are comparable to those reported by the experimental collaborations. We see that, unlike the situation at LEP2, contributions from $\dgRZu$, $\dgRZd$ as allowed by EWPD can be at a similar level as, and even dominant over those from anomalous TGCs being considered (effects of other anomalous couplings are very small and not shown).
The issue becomes more severe if the flavor universality assumption is relaxed, as vertex corrections are even less constrained in that case \cite{flavorful}. Therefore, interpreting LHC $WW$ data as a measurement of TGCs while neglecting these additional, potentially more important effects limits the generality of the results. A consistent global fit should include effects of $\dgRZu$, $\dgRZd$ along with those of $\dgz, \dka, \la$ when considering such data, and interpret the latter as constraining this extended parameter space.

As a side remark, we note that large contributions from $\dgRZu$, $\dgRZd$ are dominated by new physics amplitude squared terms rather than interference with the SM. The same is true for anomalous TGCs \cite{TGCLHC}. 
Generically, it is difficult for LHC data to be sensitive to interference terms due to limited precision. Yet, consistent interpretation of dimension-six SMEFT constraints can be made in some restricted contexts, in particular strongly-coupled scenarios, where dimension-eight operators' contributions are expected to be subdominant by power counting \cite{EFTvalidity}. 

\vspace{3pt}

{\it Toward a high energy picture of Standard Model deviations.}---The reason for the different conclusions regarding LEP2 and LHC is twofold. First, $Z$ couplings to quarks are less constrained than those to leptons; even a nonzero $\dgRZd$ is favored due to the $Z\to b\bar b$ forward-backward asymmetry anomaly \cite{LEP1}. Second, as we will discuss below, some vertex corrections, even though constrained by $Z$-pole data, lead to cross section corrections relative to the SM that grow with $\hat s$. 
Their effects are thus amplified at higher energies. 
This latter aspect will persist in the future. In particular, it has been proposed to measure $e^+e^-\to W^+W^-$ at much higher energies and precisions than LEP2 to search for deviations from the SM. But whether such deviations, if established, should be interpreted as indicating anomalous TGCs will crucially depend on our knowledge of the additional effects, which in turn depends on availability of precision data of other observables. We leave a detailed study to future work, but simply comment here that in the scenario where improved $Z$-pole measurements will not occur before the next $e^+e^-\to W^+W^-$ measurements (as envisioned for the ILC \cite{ILCos}), TGC interpretations will indeed be challenged by possibly large effects of $\dgLZv$, $\dgRZe$.

In fact, searches of SM deviations at the LHC and future colliders share a common theme of going to higher energy and taking advantage of the anomalous growth of cross sections. In the case of $f\bar f\to W^+W^-$, consider the high energy limit $v \ll \sqrt{\hat s} \ll \Lambda$, where
\begin{widetext}
\beqa
\frac{d\sigma_{f_L \bar f_R}}{d\cos\theta} &=& \frac{\pi\alpha^2(1-\cos^2\theta)}{4N_c m_Z^2\cw^4\sw^4} \bigl[ T^3_f(\cw^2-\sw^2) +Q_f\sw^2 \bigr] \Bigl[ -\dgLZfp -(T^3_f-Q_f\sw^2)\,\dgz +(T^3_f-Q_f)\,\frac{\sw^2}{\cw^2}\,\dka \Bigr] +\O\bigl(\hat s^{-1}\bigr) \CR
&=&  \frac{\pi\alpha^2(1-\cos^2\theta)}{8N_c m_Z^2\cw^4\sw^4} \bigl[ T^3_f(\cw^2-\sw^2) +Q_f\sw^2 \bigr] \bigl[ C_{HF}^{(1)} +2\,T^3_fC_{HF}^{(3)}\bigr] +\O\bigl(\hat s^{-1}\bigr) \,, \label{eq:dsigL}\\
\frac{d\sigma_{f_R \bar f_L}}{d\cos\theta} &=& \frac{\pi\alpha^2(1-\cos^2\theta)}{4N_c m_Z^2\cw^4\sw^2} Q_f \Bigl(-\dgRZf +Q_f\sw^2\,\dgz -Q_f\frac{\sw^2}{\cw^2}\,\dka \Bigr) +\O\bigl(\hat s^{-1}\bigr) 
=\frac{\pi\alpha^2(1-\cos^2\theta)}{8N_c m_Z^2\cw^4\sw^2} Q_f C_{Hf} +\O\bigl(\hat s^{-1}\bigr) \,.\qquad\label{eq:dsigR}
\eeqan
\end{widetext}
Here $\theta$ is the angle between the $W^-$ and $f$ momenta. Only terms linear in anomalous couplings or operator coefficients have been shown, which are sufficient 
for making our point in the following discussion. 
We comment in passing that unlike the case of LHC, quadratic terms are 
subdominant for $e^+e^-\to W^+W^-$ up to $\sqrt{s}\sim1$\,TeV, 
when values of vertex corrections consistent with EWPD and per-mil-level $\dgz$, $\dka$ are considered (contributions from $\lambda_\gamma$ do not grow with energy at linear level \cite{NonInterference,TGCLHC}, and can be dominated by quadratic terms).

The high energy behavior shown in Eqs.~\eqref{eq:dsigL} and~\eqref{eq:dsigR} can be easily understood and reproduced using the Goldstone boson equivalence theorem \cite{eewwNonlinear}, which states that scattering amplitudes involving longitudinal gauge bosons coincide with those involving the corresponding Goldstone bosons in the high energy limit. For example, $\O_{Hf}\supset i(\phi^-\partial_\mu\phi^+ -\phi^+\partial_\mu\phi^-)(\bar f\gamma^\mu f)$, with $\phi^\pm$ being the Goldstone bosons eaten by $W^\pm$, mediates $f_R\bar f_L\to\phi^+\phi^-$ via a contact interaction vertex, with an amplitude proportional to $\frac{\hat s}{\Lambda^2}$ by dimensional analysis. The corresponding amplitude $f_R\bar f_L\to W_L^+W_L^-$ (``$L$'' in $W^\pm_L$ for ``longitudinal'') thus also grows with $\hat s$, in contrast to the SM amplitude which $\sim\hat s^0$. On the other hand, $\O_{HWB}$, $\O_{HD}$ and $\O_{ll}$ do not mediate $f\bar f\to\phi^+\phi^-$ at tree level, while their contributions to $f\bar f\to W_T^+W_T^-$, either direct or via shifting input observables, necessarily involve factors of the Higgs vev and hence $\sim\frac{v^2}{\Lambda^2}$. Another interesting feature of Eq.~\eqref{eq:dsigL} is that $\dgLWF$ and $\dgLZf$ contribute via the combination $2T^3_f\dgLWF -\dgLZf=-\dgLZfp$ in the high energy limit. This can be seen from $SU(2)_L$-conjugating, schematically, $\frac{v^2}{\Lambda^2} (gZ_\mu) (\bar f'_L\gamma^\mu f'_L) \to \frac{\phi^+\phi^-}{\Lambda^2}(i\partial_\mu)(\bar f_L\gamma^\mu f_L)$.

The discussion above suggests that as precision studies are pushed to higher energies, it is useful to reorganize our thinking about SM deviations. Conventionally, the experimental precision hierarchy between pole observables and $f\bar f\to W^+W^-$ has motivated the use of anomalous couplings and the procedure of constraining first the parameters in Eq.~\eqref{eq:ACpole}, and then anomalous TGCs with the former set to zero. As higher energies $\sqrt{\hat s}\gtrsim v$ are reached, we are probing the electroweak symmetric phase where fully $SU(2)\times U(1)$-invariant effective operators are more useful to guide our thinking than anomalous couplings defined in the broken phase. In this regard, a better-motivated separation is between operators that lead to anomalous growth with energy for the cross sections under consideration vs.\ those that do not. 
This separation can be made also when quadratic terms, not shown in Eqs.~\eqref{eq:dsigL} and~\eqref{eq:dsigR}, are included. For $f\bar f\to W^+W^-$, the first set consists of $\O_{HF}^{(1,3)}$, $\O_{Hf}$, and also $\O_{3W}$ when quadratic terms are considered. Interestingly, $\O_{HF}^{(1,3)}$, $\O_{Hf}$ do not by themselves induce anomalous TGCs, but are turned on only to adjust $\delta g_{L,R}^{Zf}$ to zero in conventional TGC analyses; see Eq.~\eqref{eq:ACWarsaw}. 
Within the range of validity of the SMEFT ($\sqrt{\hat s}\ll\Lambda$), this set of operators is likely to be more accessible experimentally, leading to a different precision hierarchy than before.

\vspace{3pt}

{\it Conclusions.}---As precision measurements continue to explore higher energies in order to resolve SM deviations enhanced in this regime, our understanding of existing constraints also evolves; and so does the overall picture of the SMEFT parameter space. In particular, it should be kept in mind that EWPD will not always render $Zff$ vertex corrections completely irrelevant for other observables. Meanwhile, accessibility to various directions of the SMEFT parameter space will rely more heavily on high energy behaviors of effective operators, rather than the anomalous couplings they induce. We have illustrated this point in the case of $W$ pair production. The TGC three-parameter fit framework has been useful and convenient in past studies of SM deviations in such processes. But now it is time to go beyond this simplified parameterization, as the key assumption that additional new physics effects are well-constrained and negligible is already -- and will continue to be -- challenged by experimental progress at the high energy frontier. A consistent global SMEFT analysis should include not only anomalous TGCs, but all parameters whose effects are enhanced at high energy when fitting $W$ pair data, so that the results can be interpreted more generally. 

\vspace{3.5pt}

\begin{acknowledgments}
I am grateful to James~Wells for fruitful discussions and continuous encouragement. I also thank Laure~Berthier, Christophe~Grojean, Tevong~You and Bing~Zhou for helpful conversations. This research was supported by the DoE grant DE-SC0007859 and the Rackham Summer Award from the University of Michigan.
\end{acknowledgments}

\vspace{3.5pt}

{\it Note added.}---While this paper was being finalized, I became aware of \cite{EnergyHelps}, which also discusses opportunities of probing the SMEFT with high-energy data, but in a different context of oblique correction effects.


\bibliography{TGCrefs}


\onecolumngrid
\vspace{4pt}

\section*{Supplemental material}

\subsection*{1.~Alternative operator basis choice}

Our discussion can be equally well formulated in other dimension-six operator bases. As an instructive example, we consider the Strongly-Interacting Light Higgs (SILH) basis \cite{SILH07,SILH13,HiggsWindows}, which may be more familiar to some readers. To be precise, what we refer to as the SILH basis contains the following five operators,
\beqa
&&
\O_W = \frac{ig}{2}\bigl(H^\dagger\sigma^a (D^\mu H) -(D^\mu H^\dagger)\sigma^a H\bigr) \bigl(D^\nu W^a_{\mu\nu}\bigr), 
\quad
\O_B = \frac{ig'}{2}\bigl(H^\dagger (D^\mu H) -(D^\mu H^\dagger) H\bigr) \bigl(D^\nu B_{\mu\nu}\bigr), 
\CR
&&
\O_{HW} = ig(D^\mu H)^\dagger\sigma^a(D^\nu H)W^a_{\mu\nu}, 
\quad
\O_{HB} = ig'(D^\mu H)^\dagger(D^\nu H)B_{\mu\nu}, 
\quad
\O_T = \frac{1}{2} \bigl(H^\dagger (D^\mu H) -(D^\mu H^\dagger) H\bigr)^2 ,
\eeqan
in place of the Warsaw basis operators $[\O_{Hl}^{(3)}]_{11}$, $[\O_{Hl}^{(1)}]_{11}$, $\O_{HWB}$, $\O_{HD}$ in Eq.~\eqref{eq:OWarsaw} and $|H|^2 W^a_{\mu\nu} W^{a\mu\nu}$. We shall denote the SILH basis operator coefficients by $C'_i$ to distinguish from the Warsaw basis ones $C_i$. Assuming flavor universality and hence dropping generation indices, we show in the following a partial dictionary between the two bases:
\beqa
&&
C_{Hl}^{(3)} = \frac{1}{4}g^2 (C'_W+C'_{HW}) ,
\quad
C_{Hl}^{(1)} = -\frac{1}{4} g'^2 (C'_B+C'_{HB}) ,
\quad
C_{He} = C'_{He} -\frac{1}{2} g'^2 (C'_B+C'_{HB}) ,
\CR
&&
C_{Hq}^{(3)} = C_{Hq}^{\prime(3)} +\frac{1}{4}g^2 (C'_W+C'_{HW}) ,
\quad
C_{Hq}^{(1)} = C_{Hq}^{\prime(1)} +\frac{1}{12}g'^2 (C'_B+C'_{HB}) ,
\CR
&&
C_{Hu} = C'_{Hu} +\frac{1}{3}g'^2 (C'_B+C'_{HB}) ,
\quad
C_{Hd} = C'_{Hd} -\frac{1}{6}g'^2 (C'_B+C'_{HB}) ,
\CR
&&
C_{HWB} = -\frac{1}{4} gg' (C'_{HW}+C'_{HB}) ,
\quad
C_{HD} = -2C'_T +g'^2 (C'_B+C'_{HB}) .
\eeqa{eq:WarsawSILH}
By either direct calculation or using Eq.~\eqref{eq:WarsawSILH} in Eq.~\eqref{eq:ACWarsaw}, one obtains the anomalous couplings in the SILH basis,
\beqa
&&
\dgz = -\frac{1}{4(\cw^2-\sw^2)} \bigl[ (g^2-g'^2)(C'_{HW}+C'_W) +g'^2 (C'_W+C'_B) -2(C'_T-2\delta v') \bigr] ,
\CR
&&
\dka = -\frac{1}{4} g^2 (C'_{HW}+C'_{HB}) ,
\quad
\la = -\frac{3}{2}g\,C'_{3W},
\CR
&&
\delta_m = -\frac{1}{4(\cw^2-\sw^2)} \bigl[e^2(C'_W+C'_B) -2(\cw^2 C'_T -2\sw^2\delta v')\bigr] ,
\CR
&&
\dgLZe = \frac{1}{4(\cw^2-\sw^2)} \bigl[e^2(C'_W+C'_B) -(C'_T -2\delta v')\bigr] ,
\quad
\dgLZv = \frac{1}{4} (C'_T-2\delta v') ,
\CR
&&
\dgLZf = T^3_f [C_{HF}^{\prime(3)}]_{ij} -\frac{1}{2}[C_{HF}^{\prime(1)}]_{ij} 
-Q_f\frac{1}{4(\cw^2-\sw^2)}e^2(C'_W+C'_B)
+\frac{1}{2}\Bigl(T^3_f+Q_f\frac{\sw^2}{\cw^2-\sw^2}\Bigr) (C'_T-2\delta v') \quad (f=u,d) ,
\CR
&&
\dgRZf = -\frac{1}{2} C'_{Hf} -Q_f\frac{1}{4(\cw^2-\sw^2)}\bigl[e^2(C'_W+C'_B) -2\sw^2 (C'_T -2\delta v')\bigr] ,
\eeqan
where $\delta v'\equiv -\frac{1}{4} \bigl([C'_{ll}]_{1221}+[C'_{ll}]_{2112}\bigr)$. We see that imposing the TGC dominance assumption is more straightforward in the SILH basis --- one simply sets to zero $C'_W+C'_B$, $C'_T$, $\delta v'$, $C'_{He}$, $C_{Hq}^{\prime(3)}$, $C_{Hq}^{\prime(1)}$, $C'_{Hu}$, $C'_{Hd}$. This actually reflects an important motivation of the SILH basis, namely to easily incorporate EWPD constraints when analyzing diboson production and Higgs data. On the other hand, however, as the TGC dominance assumption is no longer supported by EWPD, and we wish to reorganize our understanding of the SMEFT parameter space by the high energy behaviors of effective operators, the Warsaw basis is presumably simpler to use. In particular, for $f\bar f\to W^+W^-$, the high energy limits of cross sections involve only $C_{HF}^{(3,1)}$, $C_{Hf}$ in the Warsaw basis at linear level [see Eqs.~\eqref{eq:dsigL} and \eqref{eq:dsigR}], while the additional combinations $C'_W+C'_{HW}$, $C'_B+C'_{HB}$ also contribute in the SILH basis, as can be inferred from Eq.~\eqref{eq:WarsawSILH}.

\subsection*{2.~Details of LHC $W$ pair production calculation}

Our numerical calculation of $W$ pair production at the LHC in the presence of anomalous couplings makes use of the \texttt{FeynRules} \cite{FR} model implementation \cite{BSMClong} in \texttt{MadGraph} \cite{MG}, interfaced with \texttt{Pythia} \cite{Pythia} and \texttt{Delphes} \cite{Delphes}. We follow the ATLAS 8\,TeV analysis \cite{ATLASWW}, and adopt the following event selection criteria:
\begin{itemize}
\item exactly one electron ($|\eta|<2.47$ excluding $1.37<|\eta|<1.52$) and one muon ($|\eta|<2.4$) with opposite charges;
\item the invariant mass of the electron and the muon $m_{e\mu} >10$\,GeV;
\item no jets with $p_T>25~\text{GeV}$ and $|\eta|<4.5$;
\item $p_T>25 (20)$\,GeV for the leading (subleading) lepton;
\item the relative missing transverse momentum $E_{T,\text{rel}}^\text{miss}\equiv E_T^\text{miss}\cdot\min\{\Delta\phi_\ell, \pi/2\} > 15$\,GeV, where $\Delta\phi_\ell$ is the azimuthal angle difference between missing transverse momentum and the nearest lepton.
\end{itemize}
Selected events are then binned by the leading lepton $p_T$ as in \cite{ATLASWW}; see Fig.~\ref{fig:LHC}. For each bin, the number of events obtained from simulation has been multiplied by a correction factor such that SM predictions match those used in \cite{ATLASWW}, which incorporate higher order corrections in SM calculations and more realistic detector simulation. Since our discussion is motivated by improvements on TGC constraints from LHC data reported in some recent analyses, we have included both new physics-SM interference and new physics amplitude squared terms in the calculation, as in those analyses. It should be kept in mind, though, that interpreting SMEFT constraints in the context of ultraviolet models requires more care if quadratic terms are important \cite{TGCLHC,EFTvalidity}.

\end{document}